\documentclass[a4paper,12pt]{revtex4}

\usepackage{latexsym}
\usepackage{graphics}
\usepackage{color}
\usepackage[latin1]{inputenc}
\usepackage{amssymb}
\usepackage{amsmath}
\usepackage{amsfonts}
\usepackage[dvips]{graphicx}

\usepackage{pdfpages}
\usepackage{epstopdf}

\begin{document}

\title{A simple laser system for atom interferometry}

\author{S. Merlet, L. Volodimer, M. Lours, F. Pereira Dos Santos
}                  

 \affiliation{LNE-SYRTE, Observatoire de Paris, LNE, CNRS, UPMC \\
61 avenue de l'Observatoire, 75014 Paris, France}

\begin{abstract} We present here a simple laser system for a laser cooled atom interferometer, where all functions (laser cooling, interferometry and detection) are realized using only two extended cavity laser diodes, amplified by a common tapered amplifier. One laser is locked by frequency modulation transfer spectroscopy, the other being phase locked with an offset frequency determined by an Field-Programmable Gate Array (FPGA) controlled Direct Digital Synthesizer (DDS), which allows for efficient and versatile tuning of the laser frequency. Raman lasers are obtained with a double pass acousto-optic modulator. We demonstrate a gravimeter using this laser system, with performances close to the state of the art.
\end{abstract}

\maketitle

\section{Introduction}

Inertial sensors based on atom interferometry have shown performances comparable or better than classical instruments, both in terms of sensitivity and accuracy \cite{Peters2001,Gustavson2000,Louchet-Chauvet2011,Hu2013}. State-of-the-art atomic sensors are in most cases bulky and weighty laboratory instruments, whereas, for most of their foreseen applications, such as space physics, oil exploration, geophysics and navigation, portable instruments are required. Efforts have been undertaken in order to develop mobile and rugged instruments \cite{Sorrentino2010,Schmidt2011a}, and operation outside the laboratory has been demonstrated \cite{Jiang2012,Stern2009,Muntinga2013,Bidel2013}. The ability of the sensors to operate outdoor in various, eventually harsh, environments puts stringent requirements on the subsystems, and especially on the laser part. For instruments based on rubidium atoms, robust laser systems have been developped using two different technological solutions: the first one based on GaAs semiconductor diode lasers emitting at 780 nm \cite{Laurent2006,Vogel2006,Schmidt2011}, the second one based on frequency doubling of lasers emitting at 1560 nm \cite{Thompson2003,Nyman2006,Carraz2009,Menoret2011}. In the first case, a careful design is required to ensure optimal mechanical and thermal stability of the laser bench, because the light propagates mainly in free space. The second solution, on the contrary, uses fiber-guided light and benefits from telecom-grade well-proven components. Whatever the solution retained, operation without major maintenance over extended periods of time will benefit from a well optimized and simple architecture, and especially from using a minimal number of active elements, such as lasers and 
optical amplifiers. For that purpose, phase modulation using electro-optical modulators has been used to generate additional frequencies, such as for repumping or for creating the two Raman beams out of a single laser \cite{Kasevich1991}. A major drawback of this solution lies in the presence of additional undesired sidebands, that create parasitic interferometers and relatively large systematic shifts \cite{Carraz2012}. 

In this paper, we describe a laser system, based on GaAs diode lasers, where all the functions required to run the interferometer are realized using only two diode lasers and one amplifier. Decisive elements allowing for simplification were the use of a single amplifier for two frequencies and a sophisticated scheme for operating laser diodes over an extended frequency range, for which normally distinct lasers have to be employed. This became possible through first, phase locking the frequency difference between the lasers \cite{Santarelli1994}, with an offset frequency tied to a FPGA-controlled agile DDS, allowing to change the laser frequency dynamically during the different stages of the measurement sequence and second, generating Raman beams thanks to a double pass acousto-optical modulator (AOM). With respect to most laser systems for atom interferometers, and similar to \cite{Cheinet2006}, the same lasers are used here for both laser cooling the atoms and operating the interferometer. A significant difference with our earlier work \cite{Cheinet2006} lies in the reduction of laser sources (2 ECDL instead of 3, one Tapered Amplifier (TA) instead of 2) at the price of a loss in versatility. In particular the Raman detuning of 400 MHz is (almost) fixed here, whereas it could be adjusted at will in a range of a few GHz in \cite{Cheinet2006}. On the other hand, the lasers are phase locked during the whole sequence, instead of being phase locked during the interferometer phase only and frequency locked using frequency to voltage conversion otherwise. The phase lock loop provides a much better stability of the laser frequencies: the bandwidth of the loop is much larger, of a few MHz instead of a few tens of kHz and no calibration is needed (by contrast with a frequency-to-voltage-converter whose scale factor can vary with the amplitude of the input signal and with temperature).

\section{Laser setup}

The optical setup is based on two linear extended cavity laser diodes (ECDL),
emitting at a wavelength of 780~nm. The diodes are based on the design described in \cite{Baillard2006}. Each cavity contains a laser diode chip (Sharp GH0781JA2C for ECDL1 and EagleYard Photonics EYP-RWE-0790-04000-0750-SOT01-0000 for ECDL2), a collimating lens, and a low-loss
interference filter as a frequency selective element. It is closed by a cat's eye, made of two converging lenses of 18 mm focal
lenghts, and a partially reflection mirror placed between them at their focal plane (R=30\% for ECDL1 and R=7\% for ECDL2). This mirror is located about 10 cm away from the laser chip. A lower reflectivity is required for stable operation of ECDL2, as the laser chip is anti-reflection coated. The laser field emitted by the diodes are linearly polarized. A slow but wide tuning of the laser frequencies is obtained by reacting on the
cylindrical piezo-electric (PZT) actuators which support the reflecting mirrors. The output power of the ECDL1 (resp. ECDL2) is 30~mW (resp. 50~mW) at 90 mA.

 \begin{figure}[h]
        \centering
       \includegraphics[angle=270,width=15 cm]{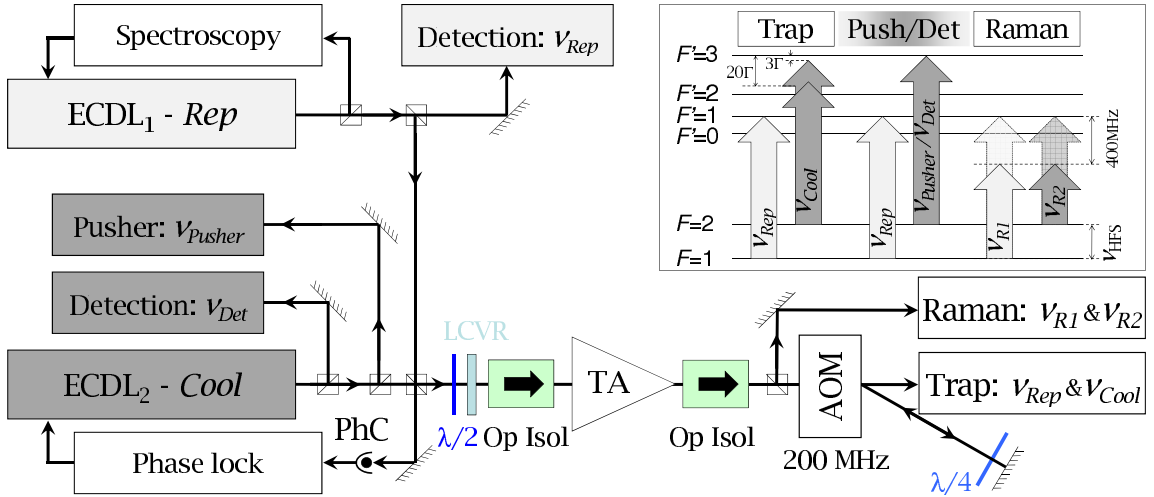}
\caption{Scheme of the laser setup, laser frequencies and energy levels involved. LCVR: Liquid Crystal Variable Retarder, AOM: Acousto-Optical Modulator, Op Isol: Optical Isolator, PhC: Photoconductor, TA: Tapered Amplifier.}
        \label{setup}
   \end{figure}

Figure \ref{setup} displays a scheme of the whole optical setup. ECDL1 is used in our cold atom interferometer setup as the repumper laser. Its output beam is splitted into three parts. A first part is used to lock the laser on the $F=1 \rightarrow F'=1$ transition of the D2 line of $^{87}$Rb by frequency modulation transfer spectroscopy \cite{Shirley1982,Ma1990} using a 5 MHz electro-optic modulator. A second part injects a fiber and is directed to the detection zone. The third and main part is mixed on a polarizing beam splitter cube with the second laser, before being sent to the TA. ECDL2 is used as the cooling laser and its output beam is also splitted in three parts: the first one injects a fiber and is directed to the detection zone, the second one injects another fiber and is used as a pusher beam, the main part is mixed with the repumper laser. One of the outputs of the polarizing beam splitter cube, where about 10\% of the output power of each laser is sent, is used to detect the beat note between the two lasers. This beat note is used to phase lock ECDL2 onto ECDL1: it is first downconverted by frequency mixing with a fixed microwave reference at 7 GHz, obtained by frequency multiplication of a 100 MHz reference signal (this 100 MHz is generated by a combination of ultralow phase noise quartz oscillators, phase-locked onto a hydrogen maser). The resulting intermediate frequency (IF) is then divided by 2 and finally compared to a DDS using a phase frequency detector, whose output error signal is used to phase lock ECDL2 onto ECDL1. This phase lock loop consists in a fast path acting onto the current of the diode laser, and a slow path acting on the PZT (the correction signal applied to the PZT is obtained by integrating the current correction signal). The architecture of this PLL is similar to the one described in \cite{Cheinet2006}. We obtain a 4 MHz bandwidth, using a simple passive lead-advance filter. 

\begin{figure}[h]
        \centering
        \includegraphics[angle=270,width=12 cm]{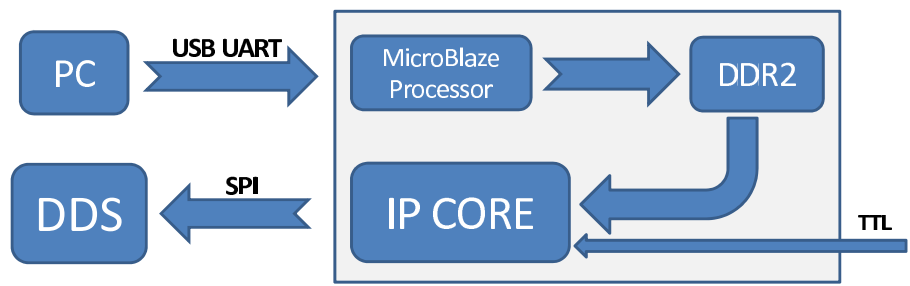}
\caption{Scheme of the control system of the DDS.}
        \label{fig:fpga2}
   \end{figure}

The DDS (AD9858 from Analog Devices) is clocked with a 1 GHz signal, phase locked onto the 100 MHz reference signal. It is controlled by an FPGA (Field-Programmable Gate Array), which allows for dynamic control of the laser frequency during the experimental sequence. The architecture of the control system of the DDS is represented in figure \ref{fig:fpga2}. 

The FPGA (Xilinx Spartan-6 LX45) is hosted by an Atlys development board, a platform for developping digital systems built around embedded processors. At first, the sequence of the successive frequencies and/or chirps, as well as their durations, that the DDS delivers in one cycle of the experiment is written in a file in the PC that controls the experiment. This file is then sent via an USB-UART port from the PC to the MicroBlaze softcore processor embedded onto the board. This processor writes the data onto the on-board DDR2 memory. An own IP (Intellectual Property) core was then developped, that reads the data from the DDR2 memory and programs the DDS via Serial Peripheral Interface (SPI) communication. The programming of the DDS is triggered by an external TTL pulse generated at the beginnning of the cycle by the PC. This ensures the synchronization of the DDS operations with the sequence of the experiment, which is generated by National Instruments boards controlled by the PC. 

\begin{figure}[h]
        \centering
        \includegraphics[width=12 cm]{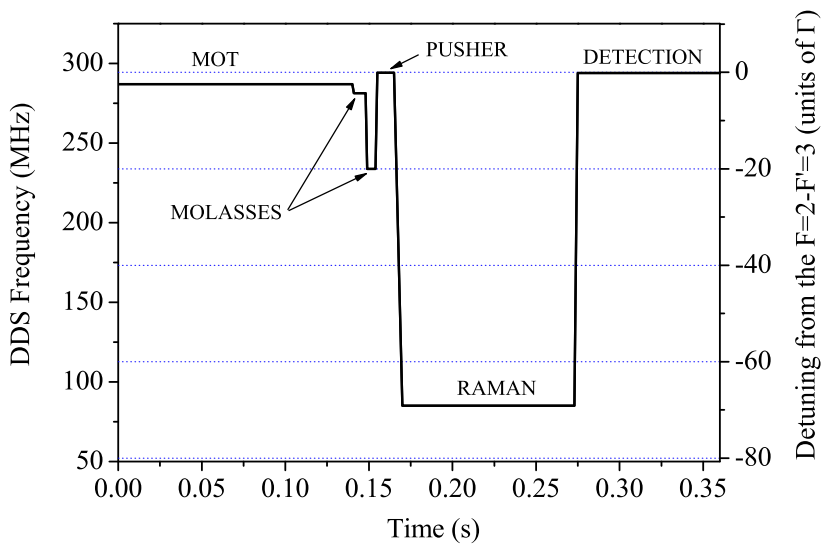}
\caption{Frequency of the DDS and corresponding detuning of ECDL2 with respect to the $F=2\rightarrow F'=3$ transition of the D2 line, as a function of time during one cycle of 360 ms.}
        \label{freq}
   \end{figure}

In practice, the DDS alternates operation at fixed frequencies with linear chirp mode of operation (for generating transient linear frequency ramps between two successive fixed frequencies). The timing resolution is limited by the reprogramming time of the DDS, on the order of $10~\mu$s, which is much shorter than the typical durations of the transient phases, on the order of a few hundreds of $\mu$s.  The frequency output of the DDS and thus of ECDL2 is displayed in figure \ref{freq}. ECDL2 is first detuned by $-3\Gamma$ with respect to the $F=2\rightarrow F'=3$ cycling transition for magneto-optical trapping, then detuned by $-5\Gamma$ for molasses, and $-20\Gamma$ for deep sub-Doppler cooling. It is then tuned back on resonance with the cycling transition to be used as a pusher, before being tuned on the $F=2\rightarrow F'=1$ transition during the interferometer sequence, and finally back to the cycling transition for detection. During this sequence, a piecewise linear voltage is also sent to the PZT controlling the cavity length of ECDL2 to pretune the laser close to the desired operating points, so that the PLL does not have to compensate for much. This ensures that the laser remains phase locked all the time, even during the transient phases, and gives better stability over time. 

The beam at the other output of the beamsplitter cube, where most of the power of each laser is sent, contains two overlapped laser beams, with different frequencies and orthogonal polarizations. Before being injected into the TA, this beam passes through an optical isolator. A set of waveplates, placed before the isolator, allows adjusting the ratio between the two laser power at the output of the isolator, and thus at the input of the TA. In particular, a liquid crystal variable retarder (Thorlabs LCC1113-B) allows for dynamical control of the input polarizations, and thus of the balance between the two beams. The retardation is controlled by the amplitude of the voltage modulation applied to the liquid crystal. The response time of this retarder being different when switching from high to low and low to high amplitudes, we choose switching from low to high amplitudes when passing from the cooling phase to the interferometer phase (for which the response time is less than 1 ms) and switching back from high to low when passing from the interferometer phase to the next cooling phase (for which the response time is 16 ms). Then, these two beams are amplified with a common TA \cite{Matsuura1998}, rated to deliver a maximum output power of 500 mW. To increase its lifetime, we operate the amplifier at a moderate current of 1.35 A and obtain 250 mW at its output. As shown in \cite{Leveque2010}, this simultaneous amplification of both lasers does not degrade the relative phase stability between the two lasers. The output of the TA then passes though a double pass acousto-optical modulator at 200 MHz. The zero-order is used for the atom trapping, as it contains repumping light from ECDL1 and cooling light from ECDL2. It injects a fiber port cluster, that splits the input into six fibered outputs. These outputs are then enlarged to 20 mm diameter beams, thanks to collimators that are directly attached to the viewports of the vacuum chamber in which the atom interferometer is realized. The doubly diffracted beam generates the pair of Raman beams used for driving stimulated Raman transitions in the interferometer. The Raman detuning is thus rather moderate, of 400 MHz only in our case. This AOM being relatively broadband, the detuning can in principle be varied in the range 300-500 MHz, at the price of a reduced diffraction efficiency. The doubly diffracted beam finally injects a polarization maintaining fiber and is sent to a collimator fixed on the top of the vacuum chamber. The collimated beam of waist 6 mm enters this vacuum chamber through a top viewport and propagates vertically down to a retroreflecting mirror located below the vacuum chamber. The overall power in the Raman beams is about 20 mW at the atoms. The value of the waist results from a compromise between reaching a  large enough Raman coupling (of order or larger than the spread of Doppler shifts in order to excite efficiently all the atoms), a sufficiently good homogeneity of the intensity across the atomic cloud (required to reach a good contrast) and a sufficiently large Rayleigh length (for minimizing systematic effects related to the Gouy phase and the curvature of the wavefronts of the laser beams \cite{Peters2001,Weiss1994} - we calculate for a waist of 6 mm a bias of -1.7 $\mu$Gal). A quarter waveplate placed on top of the retroreflecting mirror rotates by 90 degrees the polarization of the incoming beam, ensuring the proper crossed linear polarizations configuration for driving counterpropagating Raman transitions.

\section{Atom interferometer}

Our atom interferometer is realized with $^{87}$Rb atoms. We first trap directly from a vapor about $10^7$ atoms in a three dimensional magneto-optical trap (MOT) within 140 ms. The quadrupole field of the MOT is then switched off and the detuning of the ECDL1 is first tuned to $-5\Gamma$ for 8 ms, and then to $-20\Gamma$ within 1 ms. A few ms long far off detuned molasses phase, followed by an adiabatic extinction of the lasers, cools the atoms to $2~\mu$K. During this trapping and cooling phase, the variable waveplate of the bench is set such that the ratio between cooling and repumping power is about 10:1, in order to maximise the number of trapped atoms. After their release from the trap, the atoms are predominantly in the $F=2$ state. Yet, as cooling and repumping light are switched off simultaneously, about 5\% of the atoms are in the $F=1$ state. We then apply a 50 mG bias field in order to lift the degeneracy between Zeeman sensitive transitions. The atoms are then selected in the $F=1,m_F=0$ state using a microwave pulse resonant with the $F=2,m_F=0 \rightarrow F=1,m_F=0$ transition, followed by a few ms long pulse of pushing laser, realized thanks to a fast mechanical shutter. The atoms then interact with the Raman vertical lasers, in a sequence of three $\pi/2-\pi-\pi/2$ pulses, realizing a Mach Zehnder type interferometer. The duration of the $\pi/2$ (resp. $\pi$) pulse is 4 (resp. 8) $\mu$s. During the interferometer, ECDL2 is tuned on the $F=2\rightarrow F'=1$ transition, so that the frequency difference between the two beams matches exactly the hyperfine frequency difference. The Raman lasers are thus red detuned from the $F'=1$ state by 400 MHz. This detuning is large enough to ensure negligible contributions from spontaneous emission during the interferometer (we measured the contribution of spontaneous emission to the change of transition probability to be 1\% for a $\pi$ pulse). Moreover, the variable waveplate retarder is set as to obtain the ratio between the Raman laser intensities (of about 1.8) that nulls the differential light shift of the Raman lasers on the Raman resonance condition \cite{Weiss1994}. Finally, the DDS for the phase lock loop is switched to another DDS (AD 9852) during the interferometer phase, which has a better frequency resolution (48 bits instead of 32), but a lower clock frequency (300 MHz instead of 1 GHz). This resolution enables a more precise compensation of the Doppler shift of the Raman resonance condition, and thus a better control of the phase of the interferometer \cite{LeGouet2008}. The total interferometer time is 100 ms. After the interferometer, the atoms cross horizontal sheets of light that allow state dependent fluorescence detection. From the measurement of the number of atoms in each output port of the interferometer, we derive the phase of the interferometer. The total measurement cycle time is 360 ms.  

\begin{figure}[h]
        \centering
       \includegraphics[width=10 cm]{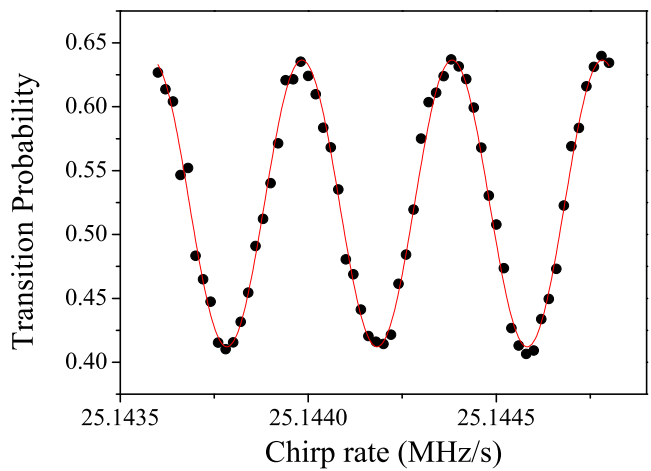}
\caption{Interferometer fringes. The phase of the interferometer is scanned by varying the chirp rate between the Raman lasers. Black dots: data, red line: sinusoidal fit to the data.}
        \label{contrast}
   \end{figure}

Figure \ref{contrast} displays the interferometer fringes measured by scanning the chirp rate between the Raman lasers. 

\begin{figure}[h]
        \centering
        \includegraphics[width=12 cm]{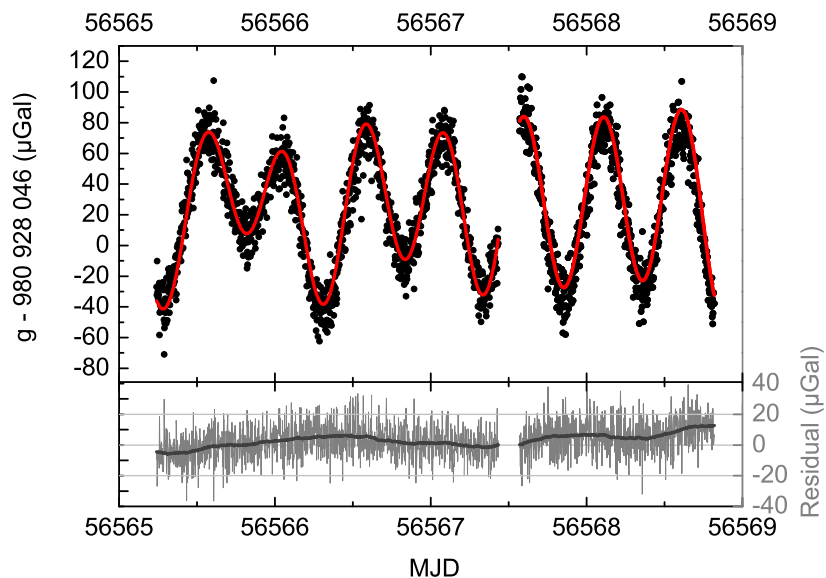}
\caption{Gravity measurement performed between September the 30th and October the 3rd 2013. The top part of the figure displays gravity measurements (black circles), averaged over 200 points (72 s) and a model for tides (red continuous line). The bottom part displays the residuals of the gravity measurements corrected from tides as a grey line. The thick dark grey line corresponds to a smoothing of the residuals (using adjacent averaging over $10000~$s).}
        \label{tides}
   \end{figure}

We obtain a maximum contrast of 22 \%, mostly limited by the velocity dispersion of the atoms ($\gtrsim 2v_r$) and the finite size of the waist of the Raman lasers. The rms phase noise is 130 mrad, which corresponds to a sensitivity to acceleration of $2\times10^{-7}$g at 1s. Post-correcting the interferometer phase from the phase shifts induced by parasitic vibration, as described in \cite{LeGouet2008}, allows reaching a sensitivity of $6\times10^{-8}$g at 1s, which is close to the level of performance of state-of-the-art instruments. We then extract the value of the gravity acceleration from the chirp rate that exactly compensates the linearly increasing Doppler frequency shift. Figure \ref{tides} displays a 3.5-days long and almost continuous measurement of gravity (black circles). The gravity measurement is performed following the measurement protocol described in \cite{Louchet-Chauvet2011}, where four different interleaved configurations are used in order to reject most of the systematics. We find a good agreement with a local model for tide variations, displayed as a red line, calculated with Etgtab v. 3.0 (Wenzel, 1996) and local tides parameters. The residuals of the gravity measurements corrected from tides are displayed at the bottom of figure \ref{tides}. We observe fluctuations in the range of $\pm10~\mu$Gal, which we attribute to fluctuations of systematic effects, such as Coriolis acceleration and wavefront distortions which are not controlled here. Though the accuracy budget of this experiment is still to be performed, the absolute value we measure is in reasonable agreement with a previous determination in the lab (performed in 2005 with a corner-cube gravimeter at a few meters distance): we find a difference as small as $8~\mu$Gal.

\section{Conclusion}
We reported here the generation of a laser system for an atom interferometer based on a simple architecture, that minimizes the number of laser sources, thanks to the use of a common tapered amplifier and of a double pass AOM to generate the Raman light beams. Some of the limitations in this design, such as the relatively low Raman power available here, could be overcome by using a more powerful tapered amplifier (2 W amplifiers are commercially available) and/or a modulator with higher efficiency (eventually in a single pass configuration). Despite the relatively modest power we use here, a large enough Raman coupling could be obtained by reducing the size of the Raman beam to a waist of 6 mm, without deleterious impact on the systematic effects. This claim is supported by the fact that the absolute gravity measurement reported here does not deviate more than $8~\mu$Gal from a former measured value. 

We then validate its functionality by operating an atom gravimeter, and demonstrate excellent performances, both in terms of sensitivity and accuracy. Indeed, though its sensitivity is currently one order of magnitude lower than the best ever demonstrated \cite{Hu2013}, it is obtained with a relatively short interferometer duration of 100 ms in a rather compact experimental setup, when compared with \cite{Hu2013}, where an atomic fountain geometry is used to reach an interferometer duration of 600 ms. In particular, its sensitivity is comparable to the one of many other such instruments \cite{Hauth2013,Altin2013,Zhou2011,Bidel2013b,Bodart2010}. We believe a more engineered version of this laser system could compete in terms of compacity, stability and cost with laser systems based on frequency doubling of telecom fibered sources. In particular, such laser systems could for instance benefit from the development of microintegrated diode lasers \cite{Luvsandamdin2013} or high power narrow linewidth DFB lasers \cite{Nguyen2012}, and from the engineering efforts undertaken for developing such laser systems for space applications \cite{Sorrentino2010,Laurent2006,Bongs2007,Grzeschik2013}.

{\bf Acknowledgments}

The authors thank Bin Wu et Emmanuel Cocher for their contribution to the development of the laser system.


\end{document}